\documentclass[twocolumn]{aastex63}
\pdfoutput=1
\usepackage{graphicx}
\usepackage{enumerate, enumitem}
\usepackage{amssymb, amsmath, amsfonts}
\usepackage{gensymb}
\usepackage{mathtools}
\usepackage{hyperref}
\usepackage{xspace}
\usepackage{xcolor}
\usepackage{ulem}

\newcommand{\masyr}{\hbox{mas\,yr$^{-1}$}}

\newcommand{\Msun}{\mbox{$M_{\sun}$}}

\newcommand{\Mjup}{\mbox{$M_{\rm Jup}$}}

\newcommand{\pmoffs}[2]{^{+ #1}_{- #2}}

\newcommand{\hipparcos}{\textit{Hipparcos}\xspace}
\newcommand{\Gaia}{\textit{Gaia}\xspace}
\newcommand{\gaia}{\textit{Gaia}\xspace}

\newcommand{\orbitcodename}{{\tt orvara}\xspace}
\newcommand{\htofcodename}{\texttt{htof}\xspace}
\newcommand{\htofversion}{\href{https://github.com/gmbrandt/HTOF/tree/0.3.1}{0.3.1}\xspace}
\newcommand{\orbitcodeversion}{githash \href{https://github.com/t-brandt/orbit3d/tree/3abd51757c58dc78db406a39503b509a929253b9}{51757c58dc78db406a39503b509a929253b9}}

\received{Nov 11, 2020}
\revised{Dec 22, 2020}
\accepted{Jan 5, 2020}

\shorttitle{Precise Dynamical Masses and Orbital Fits for $\beta$~Pic~b and $\beta$~Pic~c}
\shortauthors{Brandt et al.}
\submitjournal{AJ}

\begin{document}

\title{Precise Dynamical Masses and Orbital Fits for $\beta$~Pic~b and $\beta$~Pic~c}

\author[0000-0003-0168-3010]{G.~Mirek Brandt}
\altaffiliation{NSF Graduate Research Fellow}
\affiliation{Department of Physics, University of California, Santa Barbara, Santa Barbara, CA 93106, USA}

\author[0000-0003-2630-8073]{Timothy D.~Brandt}
\affiliation{Department of Physics, University of California, Santa Barbara, Santa Barbara, CA 93106, USA}

\author[0000-0001-9823-1445]{Trent J.~Dupuy}
\affiliation{Institute for Astronomy, University of Edinburgh, Royal Observatory, Blackford Hill, Edinburgh, EH9 3HJ, UK}

\author[0000-0002-6845-9702]{Yiting Li}
\affiliation{Department of Physics, University of California, Santa Barbara, Santa Barbara, CA 93106, USA}

\author[0000-0002-7618-6556]{Daniel Michalik}
\altaffiliation{ESA Research Fellow}
\affiliation{European Space Agency (ESA), European Space Research and Technology Centre (ESTEC), Keplerlaan 1, 2201 AZ Noordwijk, The Netherlands}

\begin{abstract}
We present a comprehensive orbital analysis to the exoplanets $\beta$~Pictoris~b and c that resolves previously reported tensions between the dynamical and evolutionary mass constraints on $\beta$~Pic~b. We use the MCMC orbit code \orbitcodename to fit fifteen years of radial velocities and relative astrometry (including recent GRAVITY measurements), absolute astrometry from \hipparcos and \gaia, and a single relative radial velocity measurement between $\beta$~Pic~A and b. We measure model-independent masses of $9.3\pmoffs{2.6}{2.5}$\,$\Mjup$ for $\beta$~Pic~b and $8.3\pm 1.0\,\Mjup$ for $\beta$~Pic~c. These masses are robust to modest changes to the input data selection. We find a well-constrained eccentricity of $0.119 \pm 0.008$ for $\beta$~Pic~b, and an eccentricity of $0.21\pmoffs{0.16}{0.09}$ for $\beta$~Pic~c, with the two orbital planes aligned to within $\sim$0.5$^\circ$. Both planets' masses are within $\sim$1\,$\sigma$ of the predictions of hot-start evolutionary models and exclude cold starts. We validate our approach on $N$-body synthetic data integrated using \texttt{REBOUND}. We show that \orbitcodename can account for three-body effects in the $\beta$~Pic system down to a level $\sim$5 times smaller than the GRAVITY uncertainties. Systematics in the masses and orbital parameters from \orbitcodename's approximate treatment of multiplanet orbits are a factor of $\sim$5 smaller than the uncertainties we derive here. Future GRAVITY observations will improve the constraints on $\beta$~Pic~c's mass and (especially) eccentricity, but improved constraints on the mass of $\beta$~Pic~b will likely require years of additional RV monitoring and improved precision from future \gaia data releases.
\end{abstract}

\keywords{---}

\section{Introduction}
$\beta$~Pictoris~b ($\beta$~Pic~b) was among the first exoplanets to be directly imaged \citep{Lagrange+etal_2010}.  Since then, it has been observed dozens of times, resulting in photometry spanning the near-infrared \citep{Quanz+Meyer+Kenworthy+etal_2010,Bonnefoy+Lagrange+Boccaletti+etal_2011,Currie+Thalmann+Matsumura+etal_2011,Bonnefoy+Boccaletti+Lagrange+etal_2013,Males+Close+Morzinski+etal_2014}, low-resolution spectroscopy \citep{Chilcote+Barman+Fitzgerald+etal_2015,Chilcote+Pueyo+DeRosa+etal_2017}, and even medium-resolution spectroscopy \citep{Snellen+etal+2014,2020AA...633A.110G}.  Part of the system's importance derives from the well-measured age of $\beta$~Pic~A.  The host star is the defining, highest-mass member of the $\beta$~Pictoris moving group \citep{BarradoyNavascues+Stauffer+Song+Caillault_1999,Zuckerman+Song+Bessell+Webb_2001}, which has concordant age determinations of $\sim$20 million years (Myr) from lithium depletion boundary measurements \citep{Binks+Jeffries_2014,Shkolnik+etal_2012}, isochrone analysis \citep{Bell+Mamajek+Naylor_2015}, and kinematic traceback \citep{Miret-Roig+Galli+Brandner+etal_2020}.

Combining this well-measured age with an independently measured mass and luminosity can constrain $\beta$~Pic~b's initial supply of thermal energy and provide clues to $\beta$~Pic b's formation mechanism \citep{Marley+Fortney+Hubickyj+etal_2007,Fortney+Marley+Saumon+etal_2008,Spiegel+Burrows_2012,Marleau+Cumming_2014}.  

Dynamical mass measurements of $\beta$~Pic~b became feasible thanks to absolute astrometry from the \hipparcos \citep{HIP_TYCHO_ESA_1997,vanLeeuwen_2007} and \gaia \citep{Gaia_General_2016,Gaia_Astrometry_2018} missions.  Since \gaia's second data release, a number of authors derived masses and orbits of $\sim$10--15~$M_{\rm Jup}$ \citep{Snellen+Brown_2018,Dupuy+Brandt+Kratter+etal_2019,Nielson+DeRosa+etal+betapicc2019}.  The picture was recently complicated and enriched by the discovery of a second companion, $\beta$~Pic~c, orbiting roughly 3~AU from the host star \citep{Lagrange_beta_pic_c, Nowak_2020_beta_pic_c_direct_detection}, interior to the $\approx$10\,AU orbit of $\beta$~Pic~b. $\beta$~Pic~c was first discovered using radial velocities alone \citep{Lagrange_beta_pic_c}, but a direct detection with GRAVITY would soon be reported. Prior to the direct detection and publication of relative astrometry of $\beta$~Pic~c, \cite{Nielson+DeRosa+etal+betapicc2019} performed a joint orbital fit to the $\beta$~Pic system. They obtained a mass of $9.4~\pm 1~\Mjup$ for $\beta$~Pic~c and $8.3\pmoffs{2.5}{2.6}~\Mjup$ for $\beta$~Pic~b (both in agreement with cooling models) but a poor constraint (roughly $\pm 13$ degrees) on the inclination of $\beta$~Pic~c \citep{Nielson+DeRosa+etal+betapicc2019} due to the lack of relative astrometry.

\cite{Nowak_2020_beta_pic_c_direct_detection} and \cite{AMLagrange2020betapicc_direct_detection} detected $\beta$~Pic~c with GRAVITY and fit a two-planet Keplerian model to the $\beta$~Pic system. When these authors adopted an uninformative prior on the mass of $\beta$~Pic~b, their best-fit dynamical mass measurements were 3.2~$\Mjup$ \citep{AMLagrange2020betapicc_direct_detection} and $5.6 \pm 1.5 \Mjup$ \citep{Nowak_2020_beta_pic_c_direct_detection}. As the authors noted, such low masses are incompatible with cooling models. Cooling models predict much higher masses that are needed to produce the observed flux (e.g., \citealt[]{Baraffe+Chabrier+Barman+etal_2003, Spiegel+Burrows_2012}).  \cite{Nowak_2020_beta_pic_c_direct_detection} ultimately adopted a prior of $15 \pm 3$~$M_{\rm Jup}$ while \cite{AMLagrange2020betapicc_direct_detection} used a prior of $14 \pm 1$~$M_{\rm Jup}$. With these priors, the posterior masses are shifted to near $\sim$10~$M_{\rm Jup}$. The necessity to use such an informative prior indicates a tension between the dynamical constraints and model predictions from spectral analyses, as noted by \cite{Nowak_2020_beta_pic_c_direct_detection}.

In this paper, we present precise masses and orbits of $\beta$~Pic~b and $\beta$~Pic~c without the need for informative priors on the planets' masses. Our inferred masses are compatible with a range of cooling model predictions and incorporate the new GRAVITY relative astrometry. We structure the paper as follows. In Section \ref{sec:data_and_fitting}, we describe the data that we adopt and the method that we use to fit the system's orbit; we include a full validation on a synthetic data set produced using $N$-body integration. We present our results in Section \ref{sec:results}, including the masses and orbital parameters of both planets, an assessment of the relative astrometry, predicted positions of $\beta$~Pic~b and $\beta$~Pic~c, and $N$-body results.  We discuss the details and implications of our work in Section \ref{sec:discussion}. We summarize our results and conclude in Section \ref{sec:conclusions}.

\section{Data and Fitting} \label{sec:data_and_fitting}

\subsection{Data}

The available data for the $\beta$~Pic system comprise more than 15 years of radial velocities (RVs) of $\beta$~Pic~A and relative astrometry for $\beta$~Pic~b, three epochs of relative astrometry for $\beta$~Pic~c, a single RV of $\beta$~Pic~b relative to $\beta$~Pic~A, and absolute astrometry of $\beta$~Pic~A from \hipparcos and \gaia. In this section we summarize each of these.  

There are several sources and numerous measurements of relative astrometry for $\beta$~Pic~b, and three recent measurements for $\beta$~Pic~c. We use all relative astrometry, which is comprised of measurements from NICI on Gemini-South \citep{Nielsen_2014_betapic_relast}, NACO on the VLT \citep{Currie_2011_Betapic_ast,Chauvin_2012AA_beta_pic_ast}, MagAO on Magellan \citep{Nielsen_2014_betapic_relast}, GPI on Gemini South \citep{Wang_2016_Betapic_relast,Nielson+DeRosa+etal+betapicc2019}, SPHERE on the VLT \citep{Lagrange2018betapic_ast_SPHERE}, and GRAVITY on the VLT (7 measurements of $\beta$~Pic~b, 3 of c) \citep{AMLagrange2020betapicc_direct_detection,Nowak_2020_beta_pic_c_direct_detection}.  This corresponds to the Case 6 relative astrometry data set of \cite{Nielson+DeRosa+etal+betapicc2019} plus recent observations by GRAVITY.

GRAVITY measurements of $\beta$~Pic~b clustered near 2020 disagree internally by $\sim$2$\sigma$ and result in an unacceptable reduced $\chi^2$ (nearly 3) on Position Angle (PA) in the final fit (see Section \ref{sec:relative_astrometry_assessment}). We therefore inflate the errors on the seven GRAVITY measurements of $\beta$~Pic~b by a factor of two to make the PA reduced $\chi^2$ acceptable and bring the PA of the 2020 measurements into internal agreement.

We use the RVs of $\beta$~Pic~A as presented in \cite{vandal_beta_pic_GP_RV_fit_2020}, which are corrected for pulsations via a Gaussian Process. We add the five new RVs presented in \cite{AMLagrange2020betapicc_direct_detection} that are not in the data set of \cite{vandal_beta_pic_GP_RV_fit_2020}.\footnote{The RVs between \cite{vandal_beta_pic_GP_RV_fit_2020} and \cite{AMLagrange2020betapicc_direct_detection} agree within the errors for the epochs mutual to the two data sets.} We also use the single measurement of the relative RV of $\beta$~Pic~b and $\beta$~Pic~A from \cite{Snellen+etal+2014}.

We use the absolute astrometry of the \hipparcos-\gaia Catalog of Accelerations \citep[HGCA,][]{brandt_cross_cal_gaia_2018}.  These astrometric measurements adopt the \gaia DR2 parallax values as priors to all \hipparcos data.  \gaia is usually much more precise than \hipparcos, but $\beta$~Pic~A is at the saturation limit of \Gaia ($G$-band magnitude of 3.7). This strongly impacts the astrometric performance of \gaia \citep{Gaia_Astrometry_2018}. Thus, the formal parallax uncertainties of the two missions are comparable (assuming a substantial error inflation to the parallax of the \hipparcos re-reduction in line with the HGCA's inflation of proper motion errors). Because the HGCA adopts \gaia parallaxes as a prior, we take the \gaia DR2 parallax value of $50.62 \pm 0.33$~milli-arcseconds (mas) \citep{Gaia_Astrometry_2018} as our prior for the orbital fit. This value is consistent to within 1\% with the \hipparcos values \citep{HIP_TYCHO_ESA_1997,vanLeeuwen_2007}. Regardless, the precise distance to $\beta$~Pic does not drive our results. 

The HGCA argues for a factor of $\sim$2 inflation for all \gaia DR2 proper motion errors. We further inflate the HGCA \gaia DR2 proper motion errors on $\beta$~Pic by another factor of 2 (a net factor of $\sim$4 over the \gaia DR2 errors). This is due to systematics in the astrometric fit for very bright stars and is justified by the black histogram (worst 5\% of stars) in Figure 9 of \cite{brandt_cross_cal_gaia_2018}. The \gaia DR2 proper motion has a negligible impact on our results with this large error inflation.

\subsection{Orbit Code}\label{sec:code_description}

We use \orbitcodename \citep{TimOrbitFitTemporary} along with \htofcodename \citep{MirekHTOFtemporary, htof_zenodo} to fit for the motion of the $\beta$~Pic system.  \orbitcodename fits one or more Keplerian orbits to an arbitrary combination of RVs, relative, and absolute astrometry. For the present analysis, we added the ability to fit the single relative RV measurement by \cite{Snellen+etal+2014}. \orbitcodename treats the full motion of the system as a linear combination of Keplerian orbits: an orbit between $\beta$~Pic~b and the combined $\beta$~Pic~A/c system, and a second Keplerian orbit between $\beta$~Pic~A and $\beta$~Pic~c.  When computing relative astrometry between $\beta$~Pic A and $\beta$~Pic~c, \orbitcodename neglects interactions with $\beta$~Pic~b.  For relative astrometry between $\beta$~Pic A and $\beta$~Pic~b, \orbitcodename computes the displacement of $\beta$~Pic~A from its center of mass with $\beta$~Pic~c and adds this to the displacement of $\beta$~Pic~b from the center of mass of the $\beta$~Pic~A/c system. In other words, \orbitcodename only adds astrometric perturbations due to inner companions, not due to outer companions. For RVs and absolute astrometry of $\beta$~Pic~A, \orbitcodename adds the perturbations from the two Keplerian orbits.  The perturbation from planet c on the relative RV measurement is negligible.

\orbitcodename uses \htofcodename to derive positions and proper motions from synthetic epoch astrometry relative to the system's barycenter.  \htofcodename uses the known \hipparcos observation times and scan angles and the predicted \gaia observation times and scan angles\footnote{\url{https://gaia.esac.esa.int/gost/}} (with dead times removed) and solves for the best-fit position and proper motion relative to the barycenter.  \orbitcodename then compares these positions and proper motions to the equivalent values in the HGCA.

\orbitcodename marginalizes out the RV zero point, the parallax, and the barycenter proper motion.  We fit a total of 16 parameters to the system using Markov Chain Monte Carlo (MCMC) with {\tt ptemcee} \citep{Foreman-Mackey+Hogg+Lang+etal_2013,Vousden+Farr+Mandel_2016}.  These are the six Keplerian orbital elements for each of planets b and c, the mass of each companion, the mass of $\beta$~Pic~A, and a RV jitter to be added in quadrature with the RV uncertainties. We adopt uninformative priors on all parameters: uniform priors on all parameters except for RV jitter (a log-uniform prior) and inclination (a geometric prior).

\subsection{Validation}

\begin{figure}
    \centering
    \includegraphics[width=\linewidth]{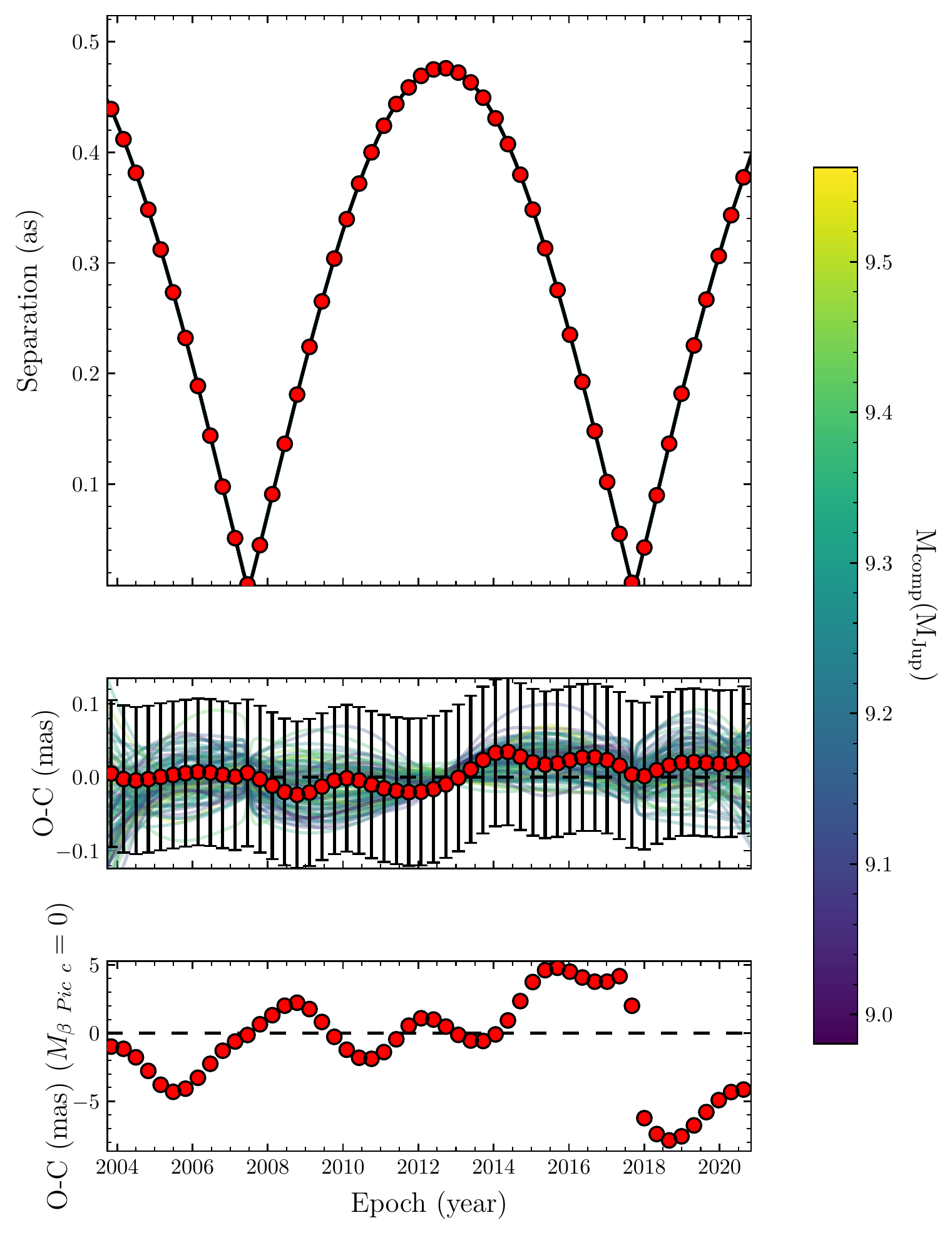}
    \caption{\orbitcodename can fit the 3-body system of $\beta$~Pic to several factors below the GRAVITY precision (assumed to be 0.1\,mas). Top panel: the observed separation for the fictitious $\beta$~Pic~b analog with evenly spaced observations with the precision of GRAVITY as presented in \cite{AMLagrange2020betapicc_direct_detection}. Black is the best fit orbit found by an \orbitcodename MCMC analysis. Middle panel: observed data minus the best fit orbit (O-C). The remaining variations are from the mutual tugs of $\beta$~Pic~c on $\beta$~Pic~b. These variations are at the level of $\sim$0.02\,mas -- a factor of 5 below the 0.1\,mas precision of the GRAVITY-like data. Bottom panel: the O-C if the approximate 3-body compensation is turned off in \orbitcodename. The synthetic data here are generated with a 9\,$\Mjup$ $\beta$~Pic~b and the best-fit mass is 9.2\,$\Mjup$.}
    \label{fig:synthetic_validation}
\end{figure}

Given the approximate treatment of the three-body problem in \orbitcodename, we test its fidelity on data integrated forward using {\tt REBOUND} \citep{rebound_2012_main}.  We initialize a 1.8 $M_\odot$ star with two planets of 9 and 8 \Mjup; we give these planets the best-fit orbital elements of $\beta$~Pic~b and c, respectively, found by \cite{AMLagrange2020betapicc_direct_detection}. We then integrate the system forward to produce synthetic RVs and relative astrometry for both companions with {\tt REBOUND}.  We take 52 measurements of relative astrometry for each planet distributed over 17 years, each of which has the 100~$\mu$as precision typical of GRAVITY \citep{2020AA...633A.110G, AMLagrange2020betapicc_direct_detection}. We fit 52 RV points, each with an uncertainty of 1~m/s. We add a single relative RV between $\beta$~Pic~b and A (the synthetic analog to the relative RV of \cite{Snellen+etal+2014}) with an uncertainty of 1 km/s. We then fit these synthetic data with \orbitcodename.

\orbitcodename is able to fit all data satisfactorily. Figure \ref{fig:synthetic_validation} shows that the unmodeled three-body effects are a factor of $\sim$5 below the level detectable by GRAVITY (see middle panel). The superposition of the two Keplerian orbits shows up in the relative astrometry of $\beta$~Pic~b, where synthetic GRAVITY observations clearly detect the orbit of $\beta$~Pic~A about its center of mass with $\beta$~Pic~c (bottom panel of Figure \ref{fig:synthetic_validation}). Unmodeled RV residuals are well below 1~m/s (the reduced $\chi^2$ of the RV fit is 0.05). We derive masses that agree well, but not perfectly, with the input masses: the derived masses of $\beta$~Pic~A and $\beta$~Pic~b are each $\sim$3\% larger than their true values. These systematics are a factor of $\sim$5 lower than the uncertainties we derive for $\beta$~Pic~c in the following section and are negligible for $\beta$~Pic~b.

\orbitcodename returns two body elements for each planet about the star. In the three body system that we initialized in \texttt{REBOUND}, the two-body input orbital elements (semi-major axis, eccentricity, etc.) cease to have a strict meaning unless a primary is specified (e.g., the barycenter or $\beta$~Pic~A). However, we still expect the recovered orbital elements to roughly be equal to those that were used as inputs. We expect the semi-major axes to be close but not exactly equal to the inputs, because, e.g., the input semi-major axis of $\beta$~Pic~b was defined relative to the barycenter of $\beta$~Pic~A and c -- yet $\beta$~Pic~b will orbit the total system barycenter during integration. Likewise, we expect the argument and time of periastron to be biased slightly. Elements like the inclination $i$ and PA of the ascending node $\Omega$ should be returned \textit{exactly} --  the 3-body interactions should not rotate the orientation of either orbit over a $\sim$20 year integration.

We find that \orbitcodename recovers $i$ and $\Omega$ exactly; with a residual less than $10^{-3}$ of a degree (nearly equal to the formal error) on both. Although unexpected, we recover the eccentricity exactly: the residual is less than $10^{-4}$ and the formal error is $2\cdot 10^{-4}$. The three elements recovered with biases follow. The argument of periastron and mean longitude at the reference epoch are recovered to within 0.2 degrees. The semi-major axes of both planets are recovered to within 0.1 A.U.


We conclude that our approximation to the three-body dynamics is more than sufficiently accurate for the $\beta$~Pic system: the biases induced in the parameters inferred from the test data are much smaller than the formal errors on the measured parameters. Our accounting of only inner companions when perturbing relative astrometry recovers the masses to within a few percent.  Figure \ref{fig:synthetic_validation} shows that a full $N$-body integration of the $\beta$~Pic system will remain unnecessary even with future GRAVITY relative astrometry.

\section{Results} \label{sec:results}
We infer masses and orbital parameters using a parallel-tempered MCMC with 15 temperatures; for each temperature we use 100 walkers with one million steps per walker.\footnote{\orbitcodename completes this million-step MCMC in roughly 4 hours on a 4 GHz AMD Ryzen desktop processor.} Our MCMC chains converged after 40,000 steps; we conservatively discard the first 250,000 as burn in and use the remainder for inference.\footnote{The chains and input data are available by request.}

We check convergence informally by confirming that we obtain the same posterior distributions, for every parameter, from any several percent portion of our chains. Next, the acceptance fraction of the coldest chain is satisfactory ($\sim$0.15). Lastly, multiple MCMC analyses starting with different, and in many cases poor, initial guesses converge to the same posterior distributions. We quantitatively confirm convergence with the Gelman-Rubin Diagnostic (GRD) \citep{Gelman+Rubin_1992_MCMCconvergence, Vivekananda_2019_MCMCconvergence}. Perfect convergence for a parameter is suggested if the GRD is 1, and a common threshold adopted for convergence is 1.1 \cite{Vivekananda_2019_MCMCconvergence}. Our chains have GRD values better than 1.0001 for all parameters, although one should keep in mind that the GRD was designed for chains with independent walkers.

\subsection{Orbital Analysis of the $\beta$~Pic System}\label{sec:orbit}

Table \ref{table:posterior_parameters} lists the six Keplerian orbital elements for both $\beta$~Pic~b and $\beta$~Pic~c, along with the other five fitted parameters.

\begin{deluxetable*}{ccccccc}
\tablewidth{0pt}
    \tablecaption{Posteriors of the $\beta$~Pic system from an \orbitcodename MCMC analysis. \label{table:posterior_parameters}}
    \tablehead{
    \colhead{Parameter} & \colhead{Prior Distribution} & \multicolumn{2}{c}{{Posteriors $\pm$1$\sigma$}}}
    \startdata
    Stellar mass & Uniform & \multicolumn{2}{c}{$1.83 \pm 0.04\,\Msun$} \\
    Parallax ($\varpi$) & $50.62 \pm 0.33$\,mas (\gaia DR2) & \multicolumn{2}{c}{$50.61 \pm 0.47$\,mas} \\
    Barycenter Proper Motions\tablenotemark{b}  & Uniform & \multicolumn{2}{c}{ $\mu_{\alpha}=4.80 \pm 0.03\,\masyr$ \; \& \; $\mu_{\delta}=83.87 \pm 0.03\,\masyr$} \\
    RV Zero Point & Uniform &
    \multicolumn{2}{c}{$33 \pm 13$ m/s} \\
    RV jitter & Log uniform over $[0, 300 {\rm \, m/s}]$ &  \multicolumn{2}{c}{$50\pm 8$\,m/s} \\
    \hline
    Parameter & Prior Distribution & Posterior on $\beta$~Pic~b $\pm$1$\sigma$ &  Posterior on $\beta$~Pic~c $\pm$1$\sigma$ \\
    \hline
    Semi-major axis ($a$) & Uniform & $10.26\pm 0.10$ A.U. & $2.738\pmoffs{0.034}{0.032}$ A.U. \\
    Eccentricity ($e$) & Uniform & $0.119\pm 0.008$ &  \tablenotemark{a}$0.21\pmoffs{0.16}{0.09}$ \\
    Inclination ($i$) & $\sin i$ (geometric) & $88.94\pm 0.02$ degrees &  $89.1\pm 0.66$ degrees \\
    PA of ascending node $(\Omega)$ & Uniform & $211.93 \pm 0.03$ degrees &  $211.1 \pmoffs{0.3}{0.2}$ degrees \\
    Mean Longitude at $t_{\rm ref}$ $(\lambda_{\rm ref})$ & Uniform & $-36.7 \pm 0.9$ degrees &  $-50\pmoffs{13}{14}$ degrees \\
    Planet Mass $(M)$ & Uniform & $9.3\pmoffs{2.6}{2.5} \Mjup$ &  $8.3\pm 1.0 \Mjup$ \\
    \hline
    Argument of Periastron $(\omega)$ & (derived quantity) & $22.6\pmoffs{2.8}{2.9}$ degrees &  $119\pmoffs{30}{7.0}$ degrees \\
    Periastron Time $(T_0)$ & (derived quantity) & $2456656\pmoffs{61}{64}$ BJD &  $2455789\pmoffs{95}{63}$ BJD \\
    Period & (derived quantity) & $8864\pmoffs{118}{113}$ days &  $1222\pmoffs{18}{17}$ days \\
     &  & $24.27\pmoffs{0.32}{0.31}$ years &  $3.346\pmoffs{0.050}{0.045}$ years \\
    \hline
    \orbitcodename Reference Epoch $(t_{\rm ref})$ & 2455197.50 BJD & \nodata & \nodata  
    \enddata
        \tablecomments{Orbital elements all refer to orbit of the companion about the barycenter. The orbital parameters for $\beta$~Pic~A about each companion are identical except $\omega_{A} = \omega + \pi$. We use $\pm$ when the posteriors are Gaussian. In the case of non-Gaussian posteriors we denote the value by median$\pmoffs{u}{l}$ where $u$ and $l$ denote the  68.3\% confidence interval about the median. The reference epoch $t_{\rm ref}$ is not a fitted parameter and has no significance within the fit itself, it is the epoch at which the Mean Longitude $(\lambda_{\rm ref})$ is evaluated.}
        \tablenotetext{a}{The posterior on the eccentricity of $\beta$~Pic~c is not Gaussian. However, eccentricities below 0.1 and above 0.7 are strongly disfavored (See Figure \ref{fig:corner_betapicc}).}
        \tablenotetext{b}{$\mu_{\alpha}$ and $\mu_{\delta}$ refer to the proper motions in right-ascension and declination, respectively.}
\end{deluxetable*}

\begin{figure*}[!ht]
    \centering
    \includegraphics[width=\linewidth]{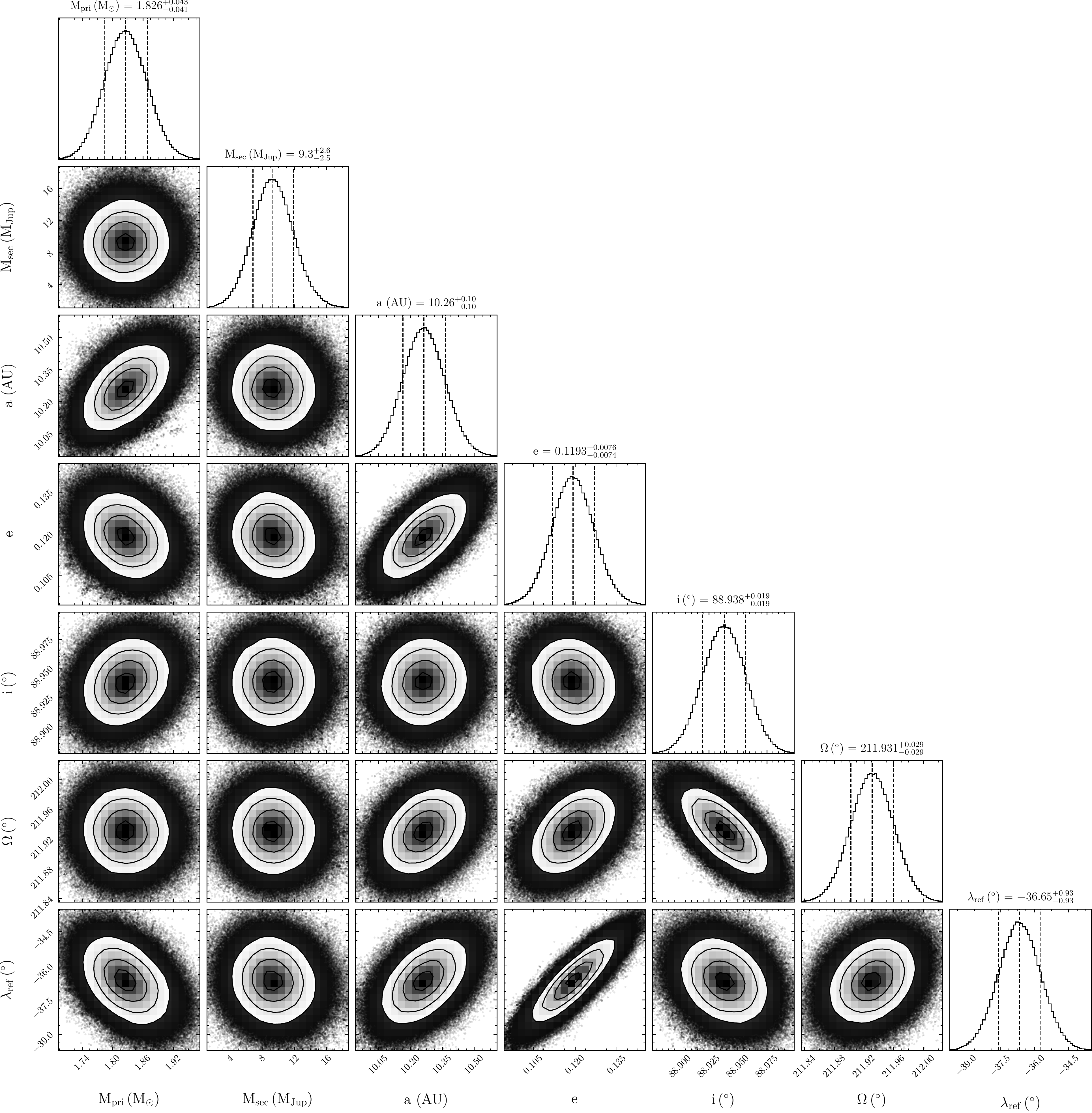}
    \caption{Best fit orbital elements for $\beta$~Pic~b from the \orbitcodename MCMC chain. Orbital elements are with respect to the star. The elements, in the same order as plotted, are: the primary mass in solar masses, $M_{\rm pri}$; the planet mass in Jupiter masses, $M_{\rm sec}$; the semi-major axis in A.U., $a$; the eccentricity, $e$; the inclination in degrees, $i$; the PA of the ascending node in degrees, $\Omega$; and the mean longitude at the reference epoch (2455197.50 BJD) in degrees, $\lambda_{\rm ref}$. In the 1D histograms, the vertical-dashed lines about the center dashed lines give the 16\% and 84\% quantiles around the median. In the 2d histograms, the contours give the 1-$\sigma$, 2-$\sigma$, and 3-$\sigma$ levels.
    }
    \label{fig:corner_betapicb}
\end{figure*}

\begin{figure*}[!ht]
    \centering
    \includegraphics[width=\linewidth]{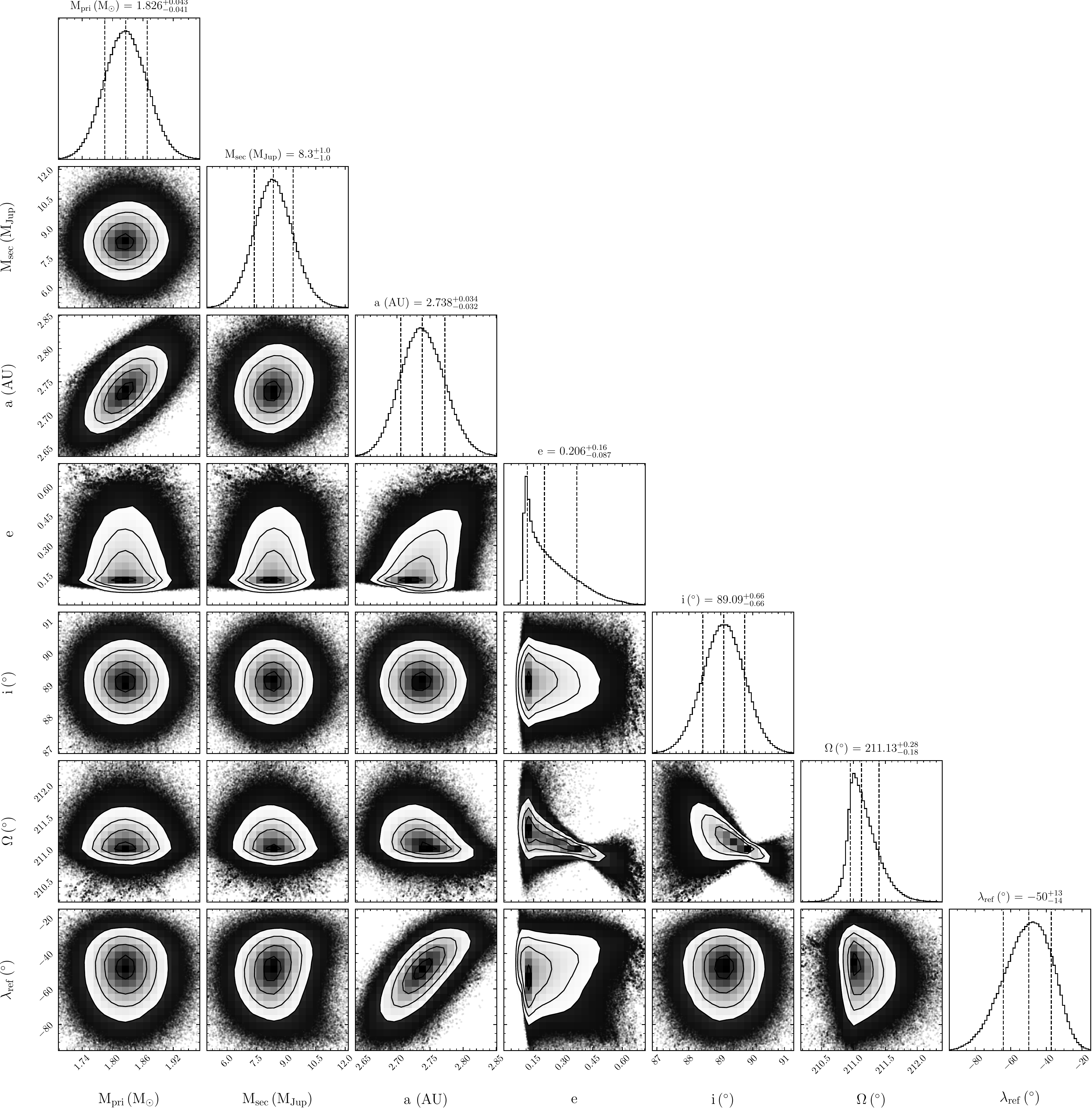}
    \caption{Best fit orbital elements for $\beta$~Pic~c. See Figure \ref{fig:corner_betapicb} for the description.}
    \label{fig:corner_betapicc}
\end{figure*}

Every fitted element of $\beta$~Pic~b results in a nearly Gaussian posterior (see Figure \ref{fig:corner_betapicb}). The elements of $\beta$~Pic~c are also well-constrained except for eccentricity and the mean longitude at the reference epoch $\lambda_{\rm ref}$. The mean longitude at the reference epoch is poorly constrained because of the poor constraint on the eccentricity, which results from having only three relative astrometric measurements closely spaced in time. We show the variances and covariances between the fitted parameters in Figure \ref{fig:corner_betapicc} for $\beta$~Pic~c as a corner plot. There is a modest covariance between semi-major axis and eccentricity resulting from the short time baseline of relative astrometry on $\beta$~Pic~c.

The best-fit orbit and nearby (in parameter space) orbits agree well with all data: the pulsation-corrected RVs, the \cite{Snellen+etal+2014} relative RV, the relative astrometry from VLT/NACO, Gemini-South/NICI, Magellan/MagAO, Gemini-South/GPI, and GRAVITY, and absolute astrometry from the HGCA.
 
\begin{figure*}[!ht]
    \centering
    \includegraphics[width=\linewidth]{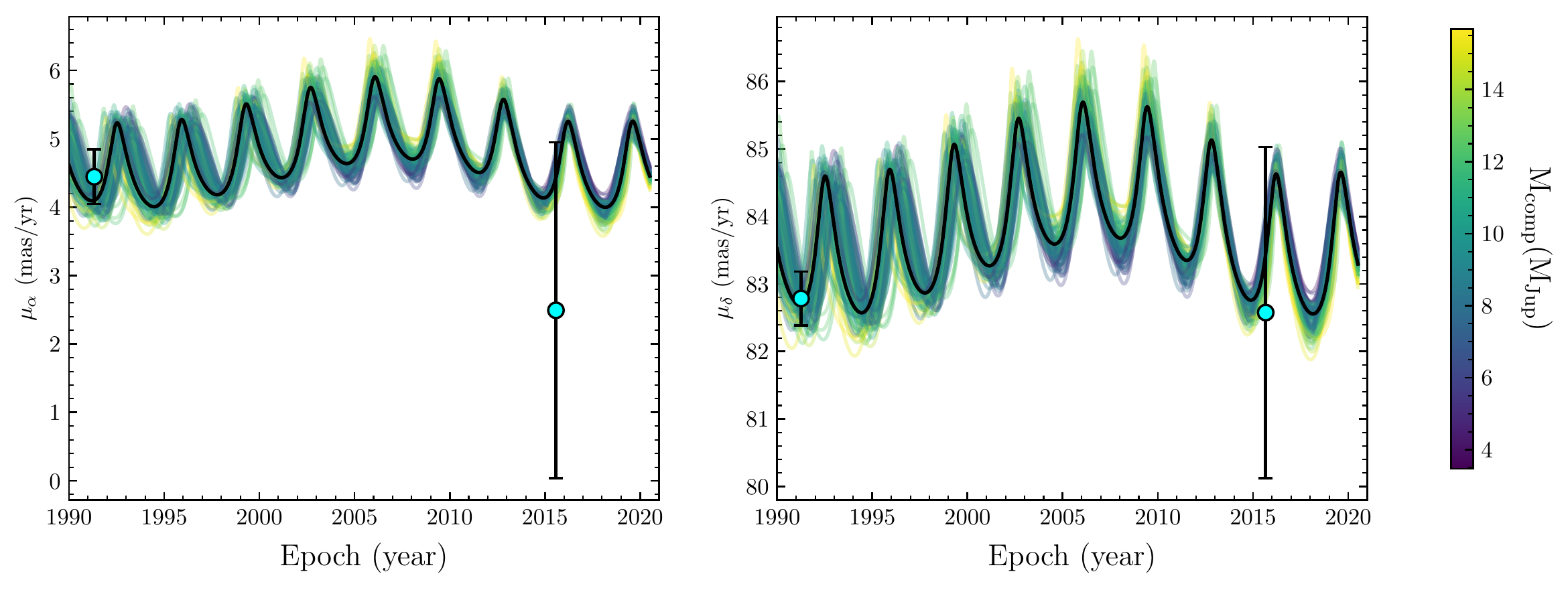}
    \caption{Model proper motions compared to the calibrated \hipparcos (dot at 1991.25) and \gaia proper motions (dot near 2015) from the HGCA. The \gaia DR2 proper motion uncertainty has been inflated by an extra factor of 2, as in \cite{Dupuy+Brandt+Kratter+etal_2019}, to account for additional uncertainties with stars as bright as $\beta$~Pic \citep[see Figure 9 of][]{brandt_cross_cal_gaia_2018}. The best fit orbit is shown in black. A random sampling of other orbits from the MCMC chain are shown and are color coded by the mass of $\beta$~Pic~b.}
    \label{fig:proper_motion}
\end{figure*}

Figure \ref{fig:proper_motion} shows the agreement between the calibrated \hipparcos and \gaia proper motions from \cite{brandt_cross_cal_gaia_2018} and the best fit orbit. The sum of the $\chi^2$ of the fits to both proper motions is very good (nearly 1, see Table \ref{table:goodness_of_orbit_fit}).  There are six measurements, but the unknown barycenter proper motion removes two degrees of freedom. The reflex motion of $\beta$~Pic~c with a period of $\sim$three years is clearly seen, as well as the long term oscillation from the $\sim$24 year orbit of $\beta$~Pic~b. Here the constraining power of the \hipparcos proper motion is visible: the \hipparcos proper motion is much more precise than that of \gaia DR2 for $\beta$~Pic~b and can exert a sizable tug on the mass and mass uncertainty of $\beta$~Pic~b.

\begin{figure*}[!ht]
    \centering
    \includegraphics[height=0.445\textwidth]{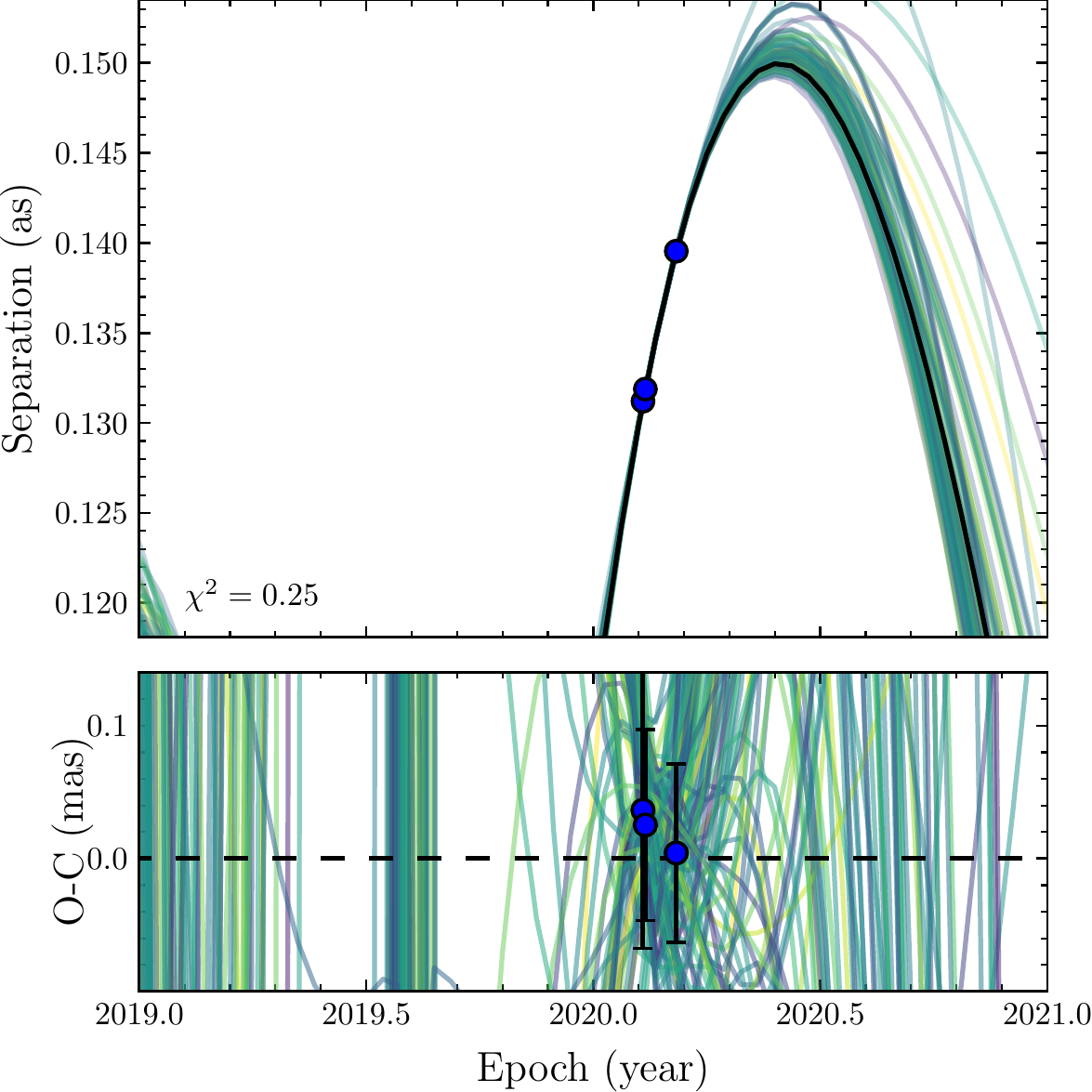}
    \includegraphics[height=0.445\textwidth]{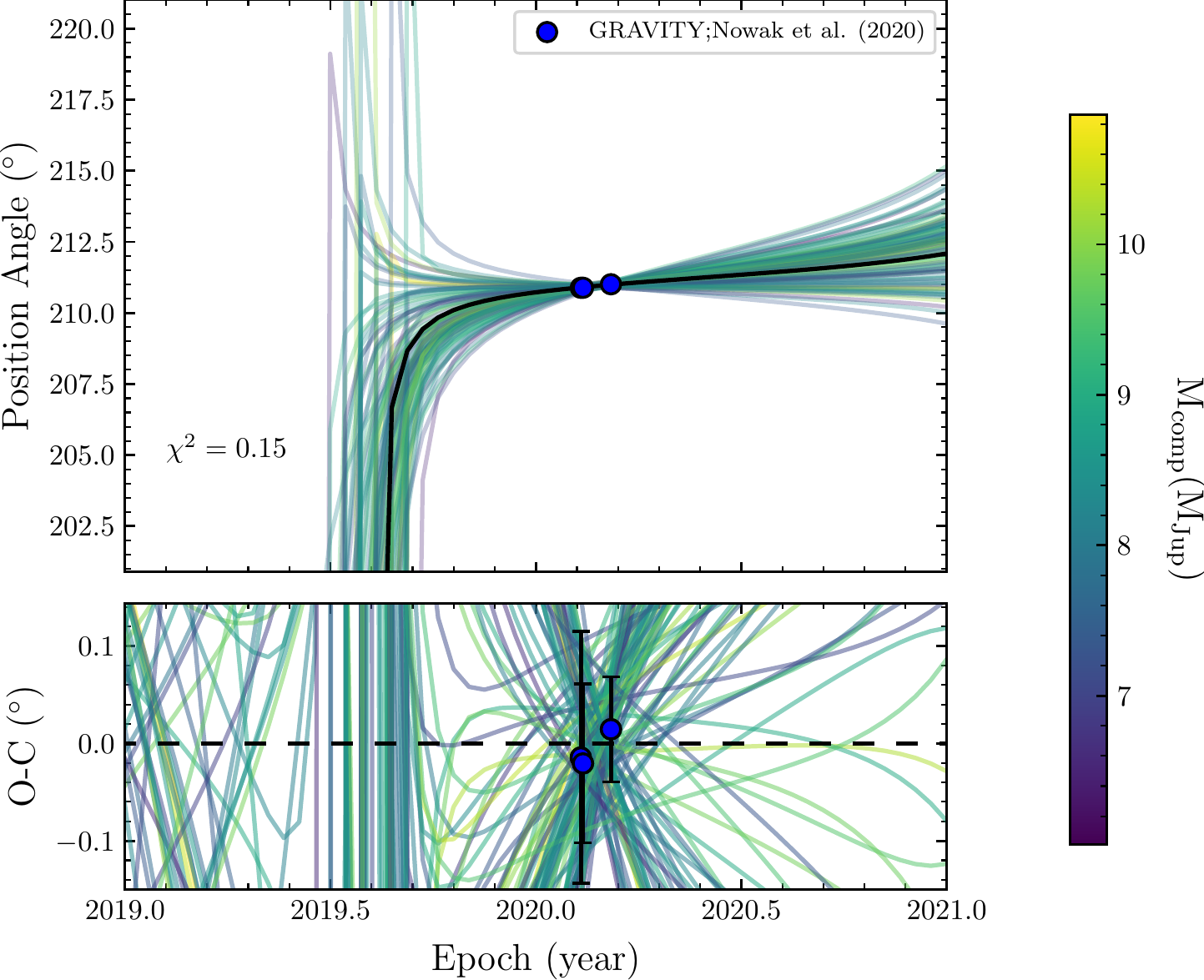}
    \caption{Left: relative separation of $\beta$~Pic~c.  Right: PA of $\beta$~Pic~c. All three data points are from GRAVITY \citep{Nowak_2020_beta_pic_c_direct_detection} and are not error inflated. The best fit orbit is shown in black.  A random sampling of other orbits from the MCMC chain are shown and are color coded by the mass of $\beta$~Pic~c.}
    \label{fig:relsep_pa_beta_picc}
\end{figure*}

\begin{figure*}[!ht]
    \centering
    \includegraphics[height=0.7\textwidth]{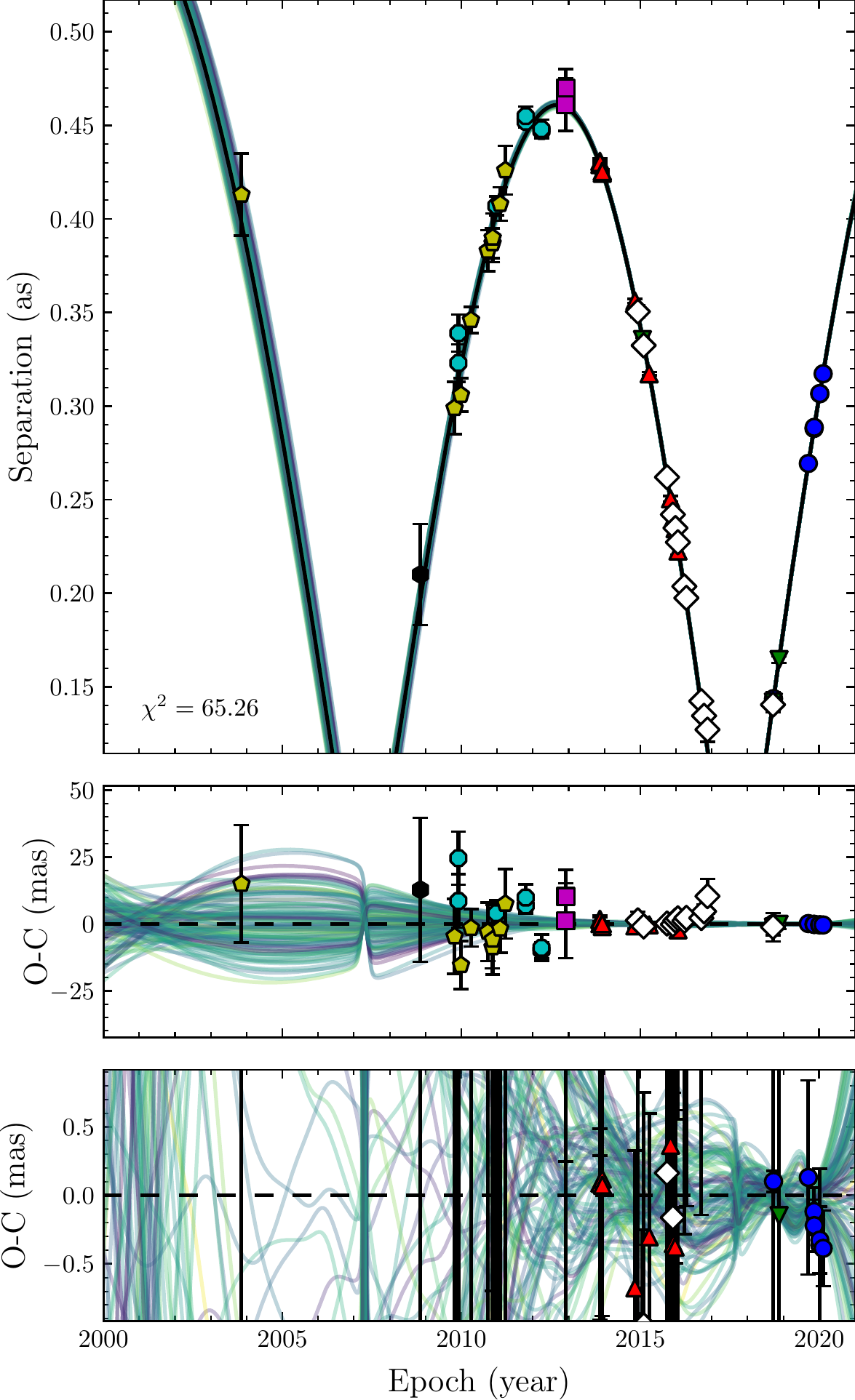} \hskip 0.1 truein
    \includegraphics[height=0.7\textwidth]{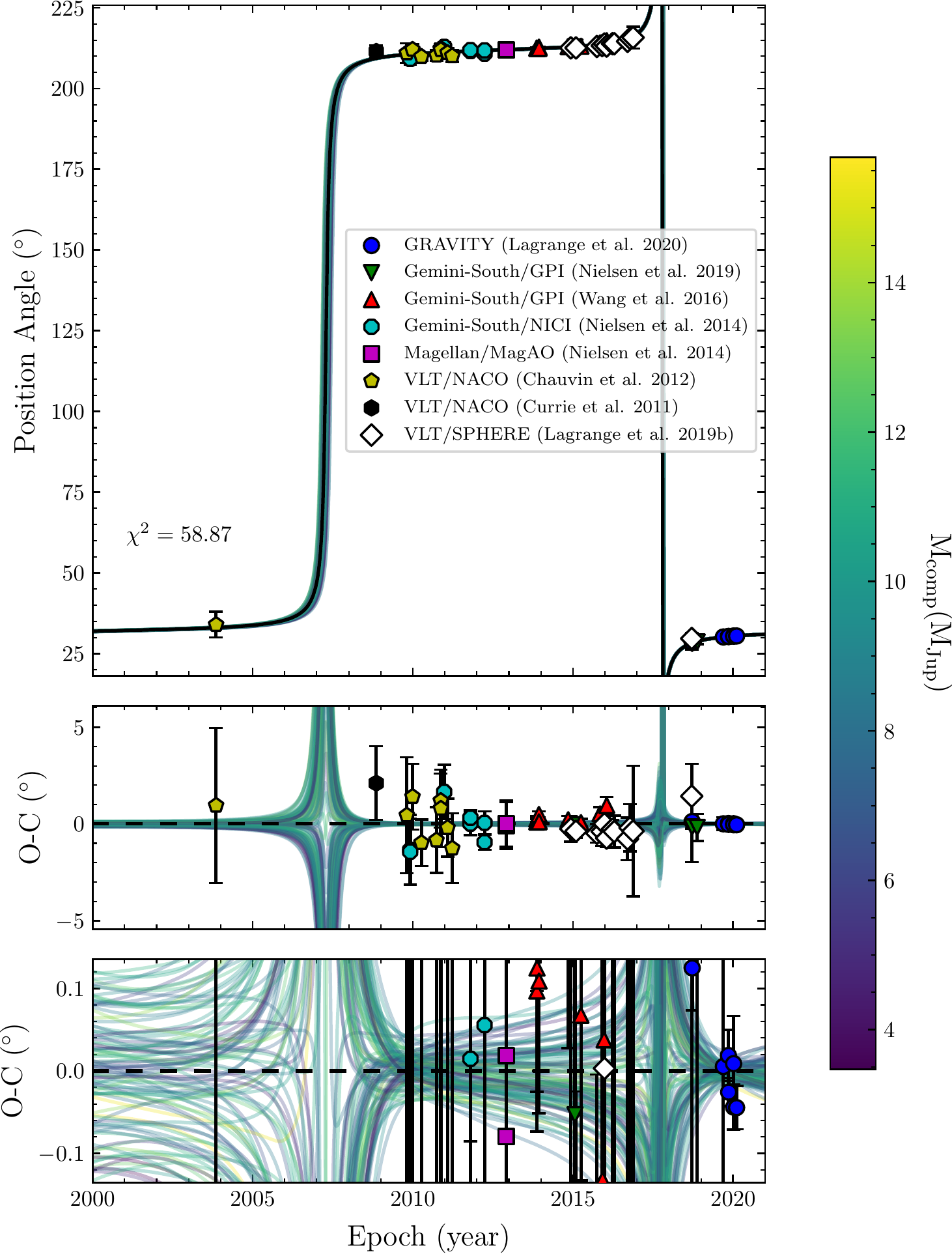}
    \caption{Left: relative separation of $\beta$~Pic~b.  Right: PA of $\beta$~Pic~b. The GRAVITY errors have been inflated by a factor of two to make the reduced $\chi^2$ of the fit acceptable. A random sampling of orbits from other MCMC steps are shown and are color coded by the mass of $\beta$~Pic~b. The best fit orbit is shown in black.}
    \label{fig:relsep_pa_beta_picb}
\end{figure*}

\begin{figure}[!ht]
    \centering
    \includegraphics[width=\linewidth]{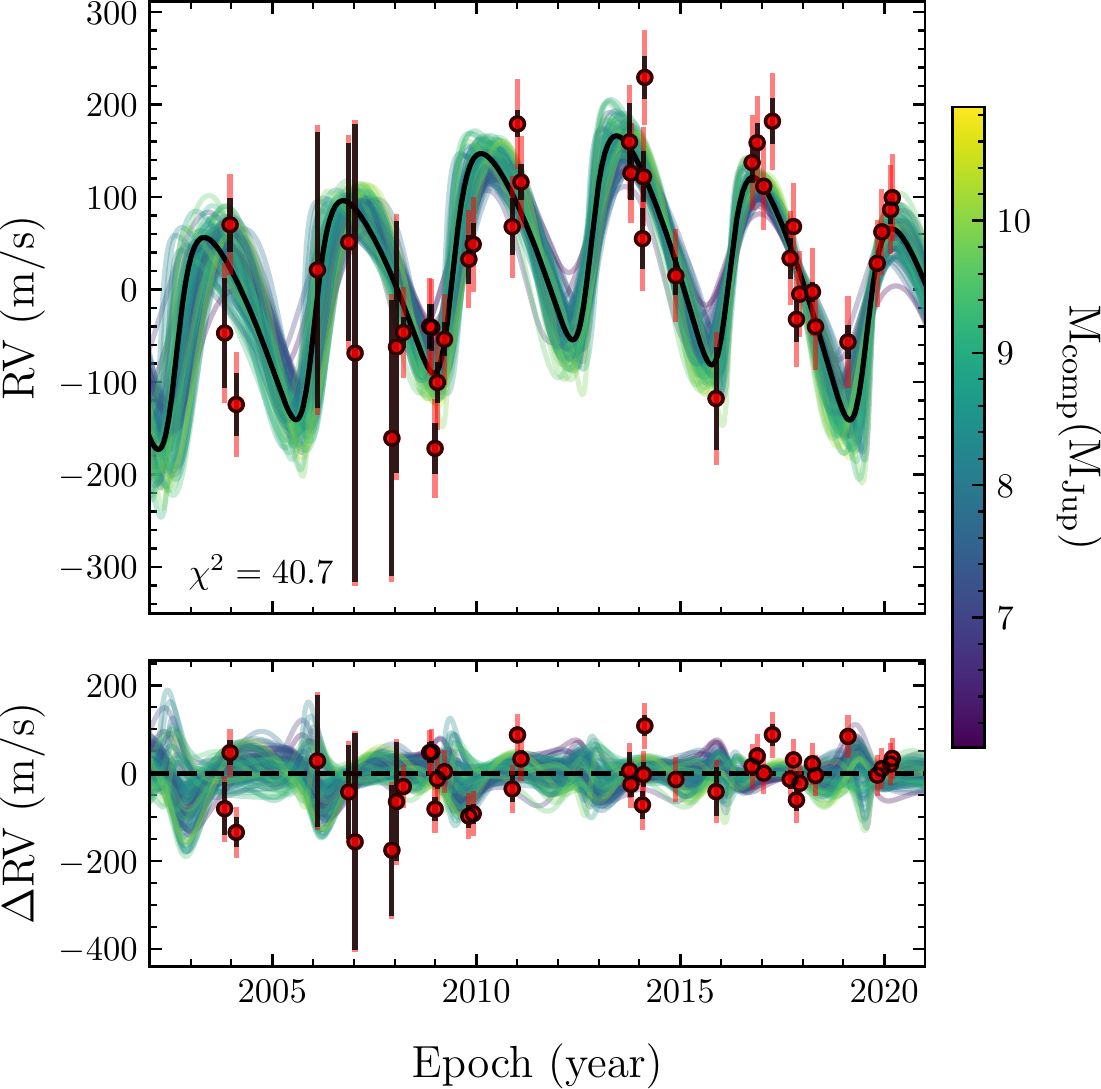}
    \caption{The best fit orbit (black) agrees well with the observed $\beta$~Pic pulsation-corrected RVs. $\beta$~Pic~c has an eccentricity of $e=0.30$ in the best fit orbit while b has $e=0.120$. Top panel: The observed RVs overplot with the best fit orbit and a random sampling of other orbits from the MCMC chain. Bottom panel: The RV residuals with respect to the best fit orbit. Both panels: The random sampling of other orbits from the MCMC chain are color coded by the mass of $\beta$~Pic~c. RVs are from \cite{vandal_beta_pic_GP_RV_fit_2020} with the most recent 5 points from \cite{AMLagrange2020betapicc_direct_detection}. The black error bars give the observed errors reported by \cite{vandal_beta_pic_GP_RV_fit_2020} and \cite{AMLagrange2020betapicc_direct_detection}. The red error bars include the best fit jitter of $\sim$50m/s added in quadrature to the observed errors.}
    \label{fig:radial_velocity_fit}
\end{figure}

Figures \ref{fig:relsep_pa_beta_picc} and \ref{fig:relsep_pa_beta_picb} show the agreement in relative separation and PA from our set of relative astrometry (Case 3 from \cite{Nielson+DeRosa+etal+betapicc2019} plus the seven GRAVITY measurements on $\beta$~Pic~b and three GRAVITY measurements on $\beta$~Pic~c). Figure \ref{fig:radial_velocity_fit} shows the agreement between the RVs from \cite{vandal_beta_pic_GP_RV_fit_2020} and \cite{AMLagrange2020betapicc_direct_detection} and the best-fit orbit. The jitter parameter found by the MCMC analysis is $50\pm 8\,{\rm m/s}$. Lower masses for $\beta$~Pic~c slightly favor lower eccentricities. The \cite{Snellen+etal+2014} relative RV $\chi^2$ is 1.7 (indicating a $\sim$1.3\,$\sigma$ residual). However, our posteriors are completely identical within rounding if we exclude the single \cite{Snellen+etal+2014} measurement.

\begin{figure*}[!ht]
    \centering
    \includegraphics[width=\linewidth]{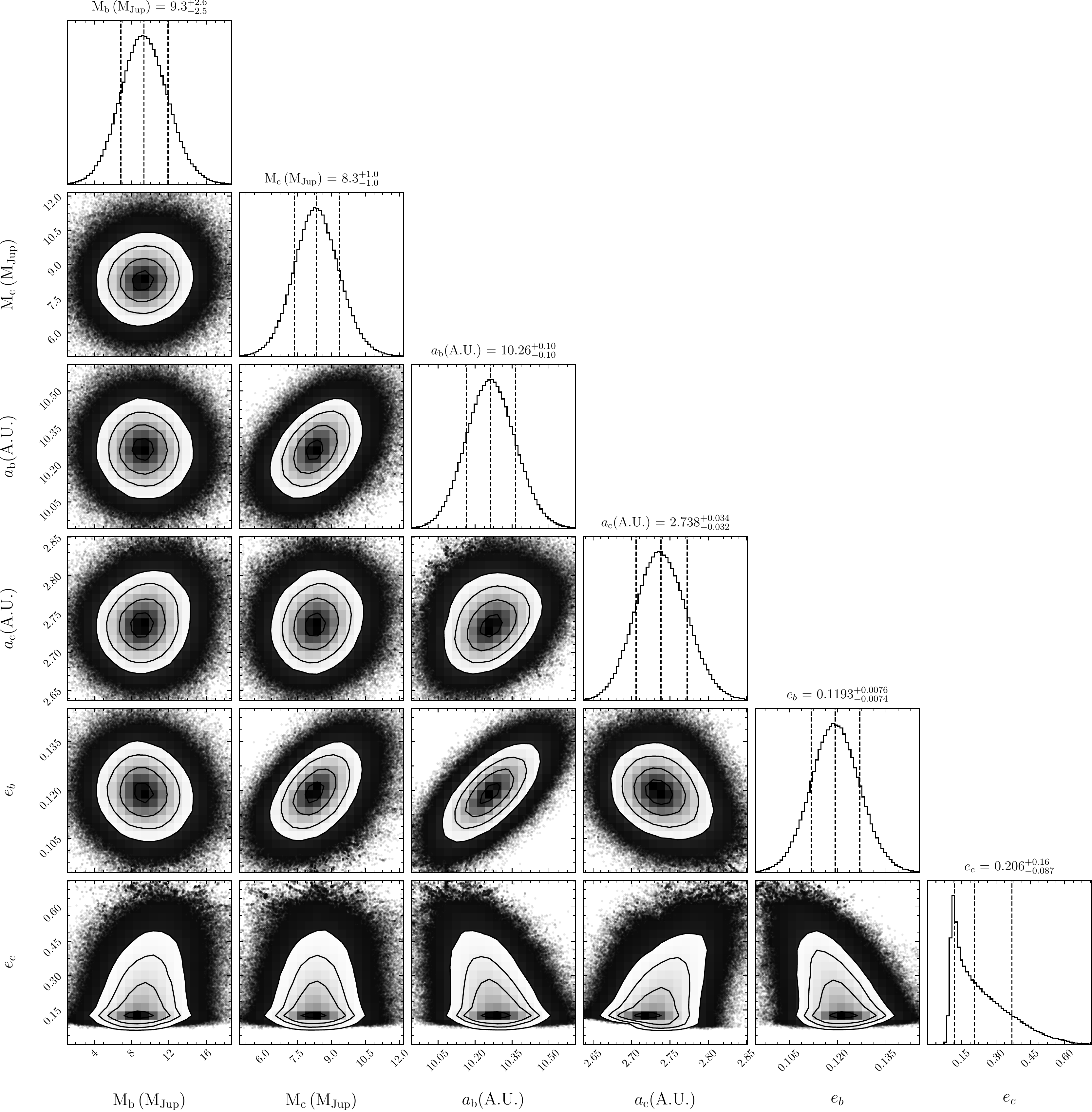}
    \caption{The masses of $\beta$~Pic~c and $\beta$~Pic~b are mostly unaffected by the eccentricity of $\beta$~Pic~c. However, the inferred eccentricity of b is moderately sensitive to the eccentricity of $\beta$~Pic~c due to 3-body interactions. We showcase here a selection of best fit orbital elements for both $\beta$~Pic~c and $\beta$~Pic~b along with the covariances between them. These are: The masses of $\beta$~Pic~b and c in Jupiter masses, $M_{\rm b}$ and $M_{\rm c}$; the semi-major axes of both planets in A.U., $a_{\rm b}$ and $a_{\rm c}$; and their eccentricities, $e_{\rm b}$ and $e_{\rm c}$. The 2d and 1d contours are described in Figure \ref{fig:corner_betapicb}.}
    \label{fig:corner_interesting_covariances}
\end{figure*}

We display an additional corner plot in Figure \ref{fig:corner_interesting_covariances} that showcases select covariances between the orbital parameters of $\beta$~Pic~b and $\beta$~Pic~c. The inferred mass of each planet is relatively insensitive to the orbital parameters of the other (see the two appropriate covariances in the left hand columns of Figure \ref{fig:corner_interesting_covariances}). In particular, the mass of $\beta$~Pic~b is nearly independent of the mass of c. However, owing to the 3-body interaction between the planets, the inferred eccentricity of $\beta$~Pic~b varies slightly with the eccentricity of the inner planet, $\beta$~Pic~c. Improved relative astrometry on $\beta$~Pic~b mildly improves constraints on the eccentricity of $\beta$~Pic~c; an identical orbital fit excluding the SPHERE relative astrometry on $\beta$~Pic~b results in a slightly worse eccentricity constraint on $\beta$~Pic~c. The inferred semi-major axis of $\beta$~Pic~c covaries modestly with $\beta$~Pic~b's eccentricity. Despite uncertainties in the eccentricity of $\beta$~Pic~c, we find that $\beta$~Pic~b and $\beta$~Pic~c are coplanar to within a half-degree at $68\%$ confidence and coplanar to within one degree at $95\%$ confidence.

We use our new constraints on the orbital parameters of $\beta$~Pic~b and c to predict their on-sky positions over the next 5 years at 15-day intervals. Tables \ref{table:betapicb_position} and \ref{table:betapicc_position} give a truncated version of the predicted positions of $\beta$~Pic~b and c. The supplementary data contain the full tables. $\beta$~Pic~c will be less than $\sim$50 mas from the star by March of 2021. $\beta$~Pic~c will re-emerge (once again being further than $\sim$50 mas from the star) in October of 2021. Our predicted positions from our orbit analysis localize both $\beta$~Pic~b and c to within $\pm~40$~mas, which is well within the fiber field of GRAVITY \citep{Nowak_2020_beta_pic_c_direct_detection}, at any point over the next 5 years.

\begin{deluxetable}{cccccccc}
\tablewidth{0pt}
    \tablecaption{Predicted positions of $\beta$~Pic~b. \label{table:betapicb_position}}
    \tablehead{
    \colhead{Date} & \colhead{$\delta$} & \colhead{$\sigma_{\delta}$} & \colhead{$\alpha$} & \colhead{$\sigma_{\alpha}$} & \colhead{$\rho_{\alpha \delta}$} & \colhead{Sep} & \colhead{$\sigma_{\rm Sep}$} \\
    \colhead{} & \colhead{mas} & \colhead{mas} & \colhead{mas} & \colhead{mas} & \colhead{} & \colhead{mas} & \colhead{mas}
    }
    \startdata
    2020-12-30 & 351.9 & 0.5 & 211.6 & 0.3 & 0.951 & 410.6 & 0.1 \\
    2021-01-14 & 355.2 & 0.6 & 213.7 & 0.4 & 0.954 & 414.5 & 0.1 \\
    2021-01-29 & 358.4 & 0.6 & 215.8 & 0.4 & 0.957 & 418.4 & 0.2 \\
    2021-02-13 & 361.6 & 0.6 & 217.9 & 0.4 & 0.959 & 422.2 & 0.2 \\
    2021-02-28 & 364.7 & 0.7 & 219.9 & 0.4 & 0.961 & 425.9 & 0.2 \\
    2021-03-15 & 367.9 & 0.7 & 222.0 & 0.4 & 0.963 & 429.6 & 0.2 \\
    2021-03-30 & 370.9 & 0.8 & 224.0 & 0.5 & 0.965 & 433.3 & 0.2 \\
    2021-04-14 & 374.0 & 0.8 & 226.0 & 0.5 & 0.967 & 437.0 & 0.3 \\
    2021-04-29 & 377.0 & 0.8 & 227.9 & 0.5 & 0.969 & 440.6 & 0.3 \\
    \nodata & \nodata & \nodata & \nodata & \nodata & \nodata & \nodata & \nodata \\
    2025-12-19 & 460 & 10 & 290 & 7 & 0.999 & 540 & 10 \\
    \enddata
        \tablecomments{The offsets, and their errors ($\sigma$), from the star in right-ascension ($\alpha$), declination ($\delta$), and separation (Sep), are given in milli-arcseconds (mas). $\rho_{\alpha \delta}$ is the correlation coefficient between right-ascension and declination. A non-rounded, machine readable version of this table with all 122 epochs is available with the supplementary data online (or with the source TeX files if viewing this on ArXiv). This portion is shown here for guidance regarding its form and content.}
\end{deluxetable}

\begin{deluxetable}{cccccccc}
\tablewidth{0pt}
    \tablecaption{Predicted positions of $\beta$~Pic~c. \label{table:betapicc_position}}
    \tablehead{
    \colhead{Date} & \colhead{$\delta$} & \colhead{$\sigma_{\delta}$} & \colhead{$\alpha$} & \colhead{$\sigma_{\alpha}$} & \colhead{$\rho_{\alpha \delta}$} & \colhead{Sep} & \colhead{$\sigma_{\rm Sep}$}
    \\
    \colhead{} & \colhead{mas} & \colhead{mas} & \colhead{mas} & \colhead{mas} & \colhead{} & \colhead{mas} & \colhead{mas}
    }
    \startdata
    2020-12-30 & $-$82 & 8 & $-$52 & 5 & 0.911 & 97 & 5 \\
    2021-01-14 & $-$76 & 9 & $-$49 & 6 & 0.927 & 90 & 5 \\
    2021-01-29 & $-$70 & 10 & $-$45 & 6 & 0.939 & 83 & 4 \\
    2021-02-13 & $-$60 & 10 & $-$41 & 7 & 0.950 & 75 & 2 \\
    2021-02-28 & $-$60 & 10 & $-$37 & 8 & 0.958 & 68 & 3 \\
    2021-03-15 & $-$50 & 10 & $-$33 & 9 & 0.965 & 60 & 8 \\
    2021-03-30 & $-$40 & 10 & $-$29 & 9 & 0.970 & 50 & 10 \\
    2021-04-14 & $-$40 & 20 & $-$20 & 10 & 0.975 & 40 & 20 \\
    2021-04-29 & $-$30 & 20 & $-$20 & 10 & 0.978 & 30 & 10 \\
    \nodata & \nodata & \nodata & \nodata & \nodata & \nodata & \nodata & \nodata \\
    2025-12-19 & 60 & 20 & 40 & 10 & 0.992 & 80 & 20 \\
    \enddata
        \tablecomments{See the table note of Table \ref{table:betapicb_position} for a description of the columns.}
\end{deluxetable}

\subsection{Assessing Consistency of Relative Astrometry} \label{sec:relative_astrometry_assessment}

Table \ref{table:goodness_of_orbit_fit} shows quantitatively the goodness of the orbital fit in terms of $\chi^2 = \sum ({\rm data - model})^2/\sigma^2$ for PA, separation, RV, and the three proper motions. The reduced $\chi^2$ for all the astrometry (which takes into account the GRAVITY covariances between separation and PA) is 1.06. A good fit should have $\chi^2/N \approx 1$ where $N$ is the number of degrees of freedom.

\begin{deluxetable}{ccc}
\tablewidth{0pt}
    \tablecaption{The goodness of the \orbitcodename orbital fit to the various data in the $\beta$~Pic system. \label{table:goodness_of_orbit_fit}}
    \tablehead{
    \colhead{Data} & \colhead{Points ($N$)} & \colhead{$\chi^2$}
    }
    \startdata
    Separation & 56 & 65.6 \\
    PA & 56 & 59.0 \\
    All Astrometry & 112 & \tablenotemark{a}118.8 \\
    RV & 41 & 40.7 \\
    $\beta$~Pic~b -- A relative RV & 1 & 1.67 \\
    Hipparcos $\mu$ (HGCA) & 2 &  0.33 \\
    Gaia $\mu$ (HGCA) & 2 & 0.75 \\
    HGCA long baseline $\mu$ & 2 & 0.001 \\
    \enddata
        \tablecomments{The $\chi^2$ quoted here include both companions and are for the maximum likelihood orbits. The $\chi^2$ for $\mu$ includes both $\mu_{\delta}$ and $\mu_{\alpha}$. $N$ is the number of data points in the corresponding data set.}
        \tablenotetext{a}{This $\chi^2$  is slightly less than the sum of the $\chi^2$ in PA and separation because of the covariance between PA and separation in the GRAVITY observations.}
\end{deluxetable}

\cite{Nielson+DeRosa+etal+betapicc2019} argued for a systematic offset between the SPHERE relative astrometry from \cite{Lagrange2018betapic_ast_SPHERE} and the relative astrometry from Gemini-South/GPI. \cite{Nielson+DeRosa+etal+betapicc2019} investigated fitting for an offset in both separation and PA within the SPHERE data. 
The SPHERE data do appear to be systematically offset in PA relative to the best-fit orbit (See the bottom right panel of Figure \ref{fig:relsep_pa_beta_picb}). However, a fit without the 12 SPHERE observations reduces the $\chi^2$ in PA and separation by roughly the expected 12 points, suggesting that the data are consistent with the astrometric record. Moreover, the reduced $\chi^2$ including SPHERE is acceptable ($\sim$65 points of $\chi^2$ for 56 data points) and so we include SPHERE in our final analysis.

We find evidence for either an underestimate in the PA uncertainties from GPI or an offset in PA between GPI and one or more of the other astrometric data sets (see the \cite{Wang_2016_Betapic_relast} GPI data in the right panel of Figure \ref{fig:relsep_pa_beta_picb}). Removing the 15 GPI relative astrometry measurements decreases the $\chi^2$ in PA by roughly 40. Using the $\chi^2$ survival function, a change of that magnitude corresponds to roughly $2.5~\sigma$ evidence in favor of a PA offset. However, including GPI still results in an acceptable overall $\chi^2$ (See Table \ref{table:goodness_of_orbit_fit}), and so we include GPI in our final fit.

Whether or not we include one, both or neither of GPI and SPHERE, our results are nearly identical. The best fit masses on both $\beta$~Pic~b and c shift by less than $0.5~\Mjup$ between all three cases, and the confidence intervals on their masses are identical to within 5\%. This speaks to the constraining power of the GRAVITY measurements, and to the robustness of our results with respect to the details of how the relative astrometry is analyzed.

In Figure \ref{fig:relsep_pa_beta_picc}, the $\chi^2$ of the $\beta$~Pic~c fit to the relative astrometry is much less than 1 because the relative astrometry is effectively overfit: the RVs primarily constrain the mass, phase, and semimajor axis of $\beta$~Pic~c while the four remaining orbital parameters have substantial freedom to fit the three relative astrometry points (6 coordinates). By contrast, $\beta$~Pic~b is overconstrained by the data and the reduced $\chi^2$ of the fit is near 1. The right-hand side of the bottom-most panel for both separation and PA in Figure \ref{fig:relsep_pa_beta_picb} show the GRAVITY points for $\beta$~Pic~b. GRAVITY points near the same epoch (in both PA and separation) disagree by $\lesssim$1$\sigma$ after error inflation.  Without error inflation, GRAVITY observations near the same epoch disagree by $\sim$2$\sigma$ and the reduced $\chi^2$ in PA of the best fit jumps to nearly 3 for $\beta$~Pic~b. 

The three GRAVITY measurements of $\beta$~Pic~c do not have $\chi^2$ or agreement issues. We leave the errors on $\beta$~Pic~c as they are in \cite{Nowak_2020_beta_pic_c_direct_detection}. However, inflating the errors by a factor of 2 on $\beta$~Pic~c does not significantly change our results: the resulting posteriors and errors are identical except for the errors on $\beta$~Pic~c's inclination, which are doubled.

\subsection{$N$-body Simulations} \label{sec:nbody}

\begin{figure}[!ht]
    \centering
    \includegraphics[width=\linewidth]{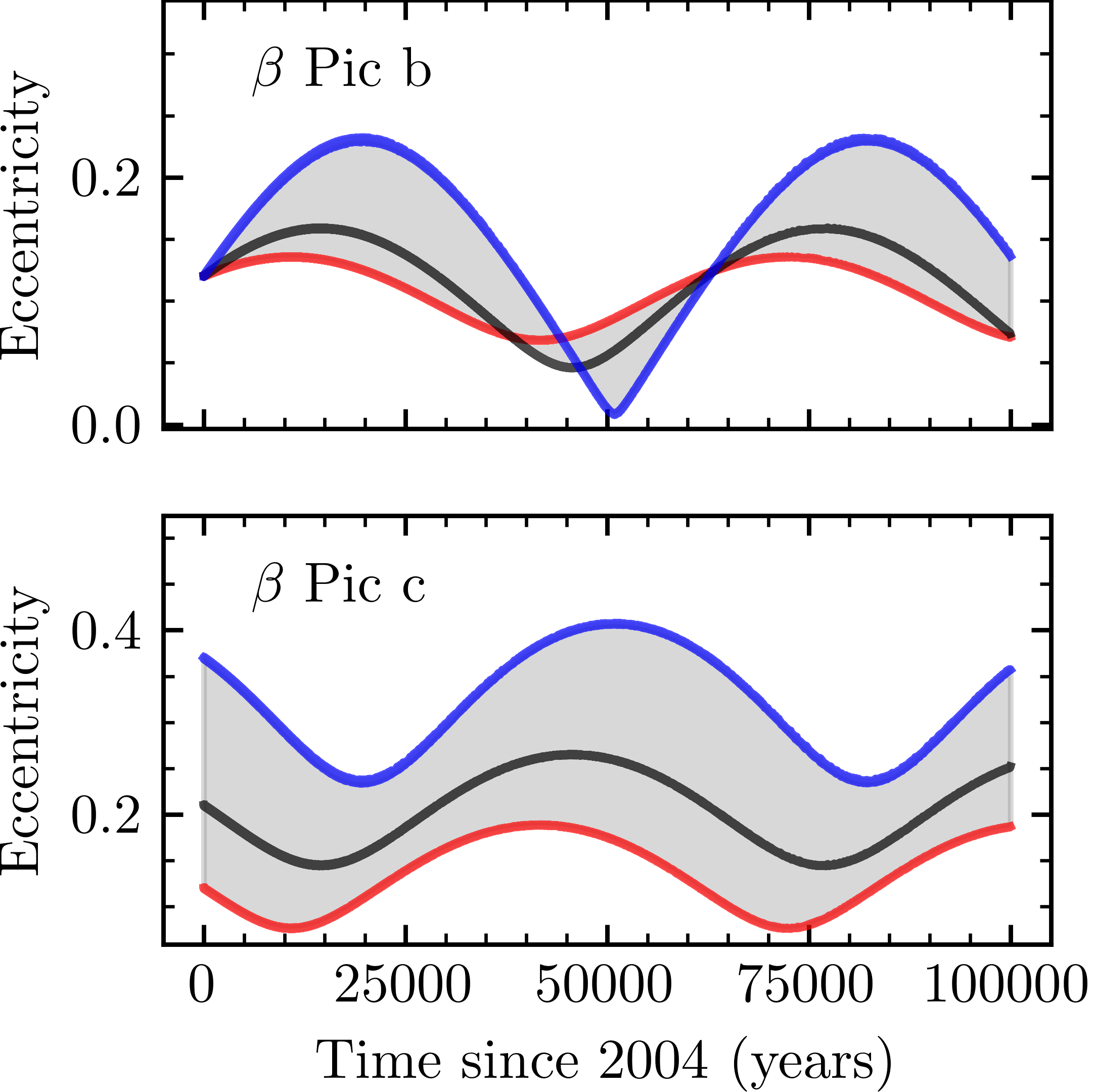}
    \caption{ 
    Top panel: the eccentricity evolution of $\beta$~Pic~b computed using {\tt REBOUND}'s ias15 integrator \citep{rebound_ias15} for the median (black, 0.21) and 68.3\% confident bounds on the eccentricity of $\beta$~Pic~c from Table \ref{table:posterior_parameters} (0.12 is red and 0.37 is blue). Bottom panel: the eccentricity evolution of $\beta$~Pic~c for its median (black) and 68.3\% confident eccentricities. The parameter space spanned by the 68.3\% confident range of eccentricities is shaded grey. The two planets exchange eccentricity over a $\sim$50,000 year cycle.}
    \label{fig:eccentricity}
\end{figure}

We expect the orbital parameters of $\beta$~Pic~b and c to vary over time due to the mutual influence between these two massive planets. The evolution of the eccentricity and orbit of $\beta$~Pic~b depends heavily on the eccentricity of $\beta$~Pic~c, which is poorly constrained. In Figure \ref{fig:eccentricity}, we show the evolution of the $\beta$~Pic system over 0.1 million years (Myr), integrated forward using the ias15 integrator of \texttt{REBOUND} \citep{rebound_2012_main, rebound_ias15}, assuming the median orbital parameters presented in Table \ref{table:posterior_parameters} for each planet. We vary the eccentricity of $\beta$~Pic~c within the posterior constraints. The grey shaded region shows how the eccentricity of $\beta$~Pic~c and $\beta$~Pic~b could evolve over the next $10^5$ years. The two planets exchange eccentricity with a period of $\sim$50,000 years. We found numerically that the system is stable and the periodic variability in Figure \ref{fig:eccentricity} repeats for at least the next 10 Myr.

\section{Discussion} \label{sec:discussion}

Our mass measurements for $\beta$~Pic~b and c agree within $1\sigma$ compared to previous work by \cite{Snellen+Brown_2018}, \cite{Dupuy+Brandt+Kratter+etal_2019}, \cite{Nielson+DeRosa+etal+betapicc2019}, and \cite{vandal_beta_pic_GP_RV_fit_2020}. Our analysis is the first to incorporate the new GRAVITY measurements with uninformative priors while obtaining masses in the expected expected range. Our error bars on the mass of $\beta$~Pic~b are larger than all but \citet{Dupuy+Brandt+Kratter+etal_2019} because, like that work, we adopt the inflated errors on the \hipparcos proper motions as recommended by \cite{brandt_cross_cal_gaia_2018}. Our mass posteriors do not change if we exclude the \cite{Snellen+etal+2014} relative RV measurement. We were unable to reproduce the $\approx$3\,$\Mjup$ and $\approx$5\,$\Mjup$ (when using a uniform prior) mass estimates from \cite{Nowak_2020_beta_pic_c_direct_detection} and \cite{AMLagrange2020betapicc_direct_detection}. Using their slightly different data set, we find $9.5\pmoffs{2.0}{1.8}\,\Mjup$ for $\beta$~Pic~b and $9.2\pmoffs{1.0}{0.8}\,\Mjup$ for $\beta$~Pic~c.

We corroborate the findings by \cite{Nielson+DeRosa+etal+betapicc2019} and \cite{Nowak_2020_beta_pic_c_direct_detection} that $\beta$~Pic~c and $\beta$~Pic~b are coplanar. \cite{Nowak_2020_beta_pic_c_direct_detection} found inclinations for $\beta$~Pic~b and c of $88.99 \pm 0.01$ degrees and $89.17 \pm 0.50$ degrees, respectively, with a Gaussian prior on the mass of $\beta$~Pic~b. We confirm these inclinations without an informative prior. We find $88.94\pm 0.02$ degrees and $89.1\pm 0.7$ degrees. 

$\beta$~Pic is surrounded by an extended debris disc and an inner disc that is slightly misaligned with respect to the primary \citep{Smith+Terrile_1984, Heap_2000_inner_debris_disc}. The extended debris disc around $\beta$~Pic is inclined at $90.0 \pm 0.1$ degrees \citep{2009ApJAhmic.Mirza_disc_inclination, 2020ApJKraus.Stefan_spinorbitbetapic}. $\beta$~Pic~b is thus misaligned by $1.06 \pm 0.11$ degrees with respect to the debris disc. Our inferred inclination for $\beta$~Pic~c slightly favors misalignment but does not exclude alignment.

\cite{Nowak_2020_beta_pic_c_direct_detection} found that the $\beta$~Pic system exhibited an oscillating eccentricity for both bodies over a timescale of $\approx5\times10^4$ years using their orbital parameter posteriors. We find variations in eccentricity over a similar timescale and confirmed numerically with \texttt{REBOUND} that the system is stable for at least the next 10\,Myr. We find that it is moderately likely to observe the current eccentricity of the system amidst all the possible eccentricities over a 10 million year timespan.


The first observational evidence that $\beta$~Pic~b has a significant, nonzero eccentricity was presented by \citet{Dupuy+Brandt+Kratter+etal_2019}. They discussed the implications of an eccentricity as high as $\approx$0.2 in the context of both single- and multi-planet scenarios; at the time $\beta$~Pic~c was not known. The scenario in which $\beta$~Pic~b formed on a circular orbit but gained eccentricity from interactions with the disk and migrated inward to its current location, with no influence from $\beta$~Pic~c, is still plausible. Such a pathway is available to any sufficiently massive planet. Given that we find that $\beta$~Pic~c is also massive ($8.3 \pm 1$\,\Mjup), it may have also opened a gap in the disk, migrating inward and acquiring eccentricity from gravitational interactions with $\beta$~Pic~b and the disk. Indeed, with two such massive planets in close proximity it is natural to expect that both should have significantly nonzero eccentricities by a system age of $\approx$20\,Myr.

\begin{figure}
    \includegraphics[width=\linewidth]{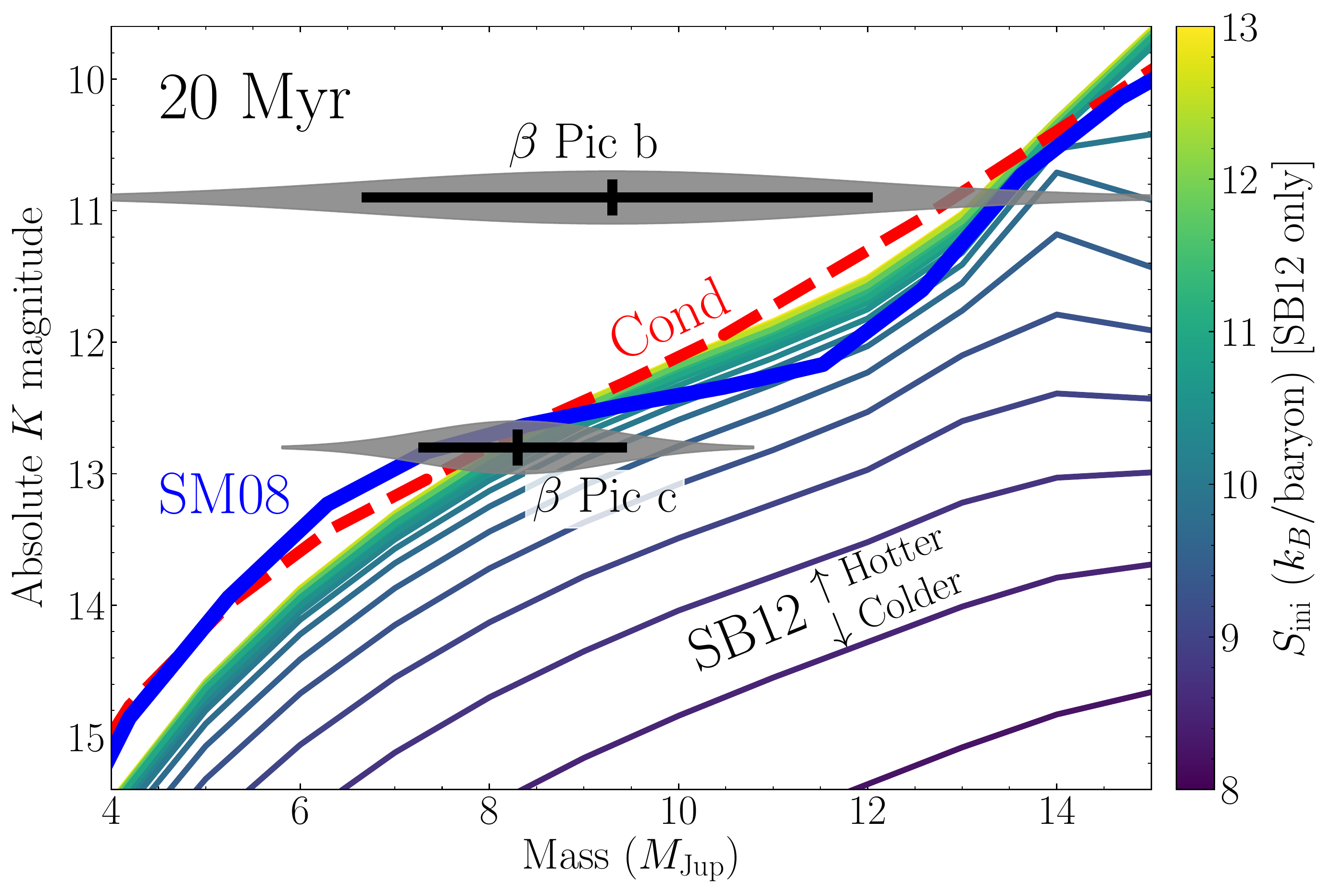} 
    \caption{Comparison of dynamical mass measurements (gray shaded regions) and observed $K$-band magnitudes \citep{Nowak_2020_beta_pic_c_direct_detection} with Cond \citep{Baraffe+Chabrier+Barman+etal_2003}, SM08 \citep{Saumon+Marley_2008}, and SB12 \citep{Spiegel+Burrows_2012} models, all at an age of 20\,Myr \citep{Binks+Jeffries_2014,Mamajek+Bell_2014,Miret-Roig+Galli+Brandner+etal_2020}.  The SM08 and SB12 models both use hybrid cloud prescriptions and adopt Solar metallicity.  The SB12 models also vary (and are color-coded by) their initial entropy.  The black lines show 1\,$\sigma$ values, while gray shaded regions show the probability density.  Our dynamical mass for $\beta$~Pic~c is consistent with all three models assuming a hot start, and rules out a very cold start.  Our dynamical mass for $\beta$~Pic~b is $\sim$1\,$\sigma$ below the prediction of the hot start models.}
    \label{fig:modelcomparison}
\end{figure}

Our masses follow from uniform priors, allowing us to independently assess the agreement of the dynamical masses with model predictions. To simplify our model comparisons, we assume an age of 20\,Myr for the system, compatible with all available age determinations for the $\beta$~Pic moving group \citep{Binks+Jeffries_2014,Mamajek+Bell_2014,Miret-Roig+Galli+Brandner+etal_2020}.
We examine the hot-start Cond \citep{Baraffe+Chabrier+Barman+etal_2003} models, the \cite{Saumon+Marley_2008} models with a hybrid cloud treatment (which we denote as SM08), and the warm-start \cite{Spiegel+Burrows_2012} models (SB12) with hybrid clouds and solar metallicity but a range of initial entropies.  We perform our comparisons in the $K$ band, as this is the only measurement available for $\beta$~Pic~c \citep{Nowak_2020_beta_pic_c_direct_detection}. We convert luminosities to $K$-band magnitudes for the SM08 models using Cond colors at the SM08 effective temperatures.

Figure \ref{fig:modelcomparison} shows our results.  We find that our dynamical mass measurement for $\beta$~Pic~c is consistent with all models except those with low initial entropies ($\lesssim 10\,k_{\rm B}/{\rm baryon}$).  Our dynamical mass for $\beta$~Pic~b is roughly 1\,$\sigma$ below the predictions of hot-start models, and rules out cold starts.  Similarly to previous work \citep{Dupuy+Brandt+Kratter+etal_2019, vandal_beta_pic_GP_RV_fit_2020}, none of the disagreements with models are significant beyond $\sim$1$\sigma$, and the precisions of the dynamical masses are insufficient to distinguish between most of the models shown.  Stronger tests of models will require significantly better precision, especially for $\beta$~Pic~b.


As Figure \ref{fig:modelcomparison} shows, reaching 0.1--0.5\,$\Mjup$ levels of precision on the mass of $\beta$~Pic~b is crucial to accurately discern between evolutionary models. The best prospect for improving the mass of $\beta$~Pic~b is long term RV monitoring over the next decade. Even drastically improved absolute astrometry (e.g., \gaia DR4) will only provide a modest improvement to the mass measurement of $\beta$~Pic~b. 
If we assume optimistically that \gaia at the end of its mission will achieve the same precision on the $G=3.7$\,mag $\beta$~Pic~A as it has on $G\approx6$\,mag stars (the brightest for which the mission was originally designed), then it would achieve a factor of $\sim$100 improvement on the proper motion of $\beta$~Pic~A.\footnote{\url{https://www.cosmos.esa.int/web/gaia/science-performance}} The uncertainty on the mass of $\beta$~Pic~b would shrink by 35\%, to $\pm 1.7 \Mjup$, if the proper motion precision is improved by a factor of 100--using the same MCMC analysis as presented here with otherwise the same data. Assuming more conservatively that \gaia reaches only a factor of 10 better precision on the proper motion of $\beta$~Pic, the uncertainty on the mass of $\beta$~Pic~b improves by 25\%.

\section{Conclusions}\label{sec:conclusions}

In this paper we have derived masses and orbits of both planets in the $\beta$~Pictoris system with uninformative priors. We validated our approach against synthetic data from a full $N$-body integration. Our masses and orbital parameters are derived from two decades of observational data. The GRAVITY data show clear evidence of the gravitational perturbations of $\beta$~Pic~c ($P=3.346^{+0.050}_{-0.045}$\,yr) on the orbit of $\beta$~Pic~b relative to A ($P=24.27\pm0.32$\,yr). The resulting model-independent masses allow us to compare the observed properties of $\beta$~Pic~b and c with predictions from models of the formation and evolution of giant planets.  We summarize our main results below.
\begin{itemize}[noitemsep]
    \item [1.] We find a mass of $9.3\pmoffs{2.6}{2.5}~\Mjup$ for $\beta$~Pic~b and $8.3 \pm 1.0\,\Mjup$ for $\beta$~Pic~c with uninformative priors all orbital parameters. The mass constraint on $\beta$~Pic~c is superior due to the RVs covering many orbital periods and due to the impact of $\beta$~Pic~c on the relative astrometry of $\beta$~Pic~b.
    \item [2.] $\beta$~Pic~b and $\beta$~Pic~c are both consistent with \cite{Spiegel+Burrows_2012} warm-start models with initial entropies of at least $10\,k_{\rm B}/{\rm baryon}$. They are also both consistent with a 20\,Myr age under the hot-start COND evolutionary tracks \citep{Baraffe+Chabrier+Barman+etal_2003} and the \cite{Saumon+Marley_2008} models using a hybrid cloud model. In all cases, consistency with models would favor a mass for $\beta$~Pic~b that is $\sim$1$\sigma$ higher than our dynamical measurement.
    \item [3.] We find an eccentricity of $0.119\pm 0.008$ for $\beta$~Pic~b and $0.21\pmoffs{0.16}{0.09}$ for c. These modest eccentricities could have been generated by interactions with the disk, or via the mutual interactions between b and c. The eccentricity and mean longitude of $\beta$~Pic~c are poorly constrained because there are only three relative astrometric observations, and these are closely spaced in time. There is a mild covariance between the eccentricity of $\beta$~Pic~b and c owing to the three-body dynamics in the system. Additional GRAVITY relative astrometry on $\beta$~Pic~c will help constrain the eccentricity of $\beta$~Pic~b and especially $\beta$~Pic~c.
    \item [4.] The mass constraint on $\beta$~Pic~b needs to be improved by a factor of $\sim$3--5 in order to more reliably constrain its age or formation conditions. Long-term RV monitoring over the coming years or decade is needed for better mass constraints on $\beta$~Pic~b. An improved proper motion from a future \gaia data release will offer up to a 35\% better constraint on the mass of $\beta$~Pic~b (assuming \gaia reaches a better precision on the brightest stars).
\end{itemize}

The new GRAVITY relative astrometry \citep{Nowak_2020_beta_pic_c_direct_detection, AMLagrange2020betapicc_direct_detection} appeared to create tension between dynamical and spectral mass constraints on $\beta$~Pic~b. Our analysis dissolves this tension and results in masses for $\beta$~Pic~b and $\beta$~Pic~c that are consistent with warm and hot start evolutionary models. Additionally, the system is dynamically interesting -- with eccentricities of both planets varying by $\sim$50\% over $10^4$--$10^5$ year timescales. The precision on the masses and eccentricities of $\beta$~Pic~b and c will improve with continued astrometric and RV monitoring. The planets around $\beta$~Pic~A will continue to provide some of the best tests of super-Jovian planet formation and evolution.

\vspace{14pt}  

\software{astropy \citep{astropy:2013, astropy:2018},
          scipy \citep{2020SciPy-NMeth},
          numpy \citep{numpy1, numpy2},
          pandas \citep{mckinney-proc-scipy-2010, reback2020pandas},
          \orbitcodename \citep{TimOrbitFitTemporary},
          \htofcodename \citep{htof_zenodo, MirekHTOFtemporary},
          REBOUND \citep{rebound_2012_main},
          corner \citep{corner},
          Jupyter 
          }

\acknowledgments
{G.~M. Brandt is supported by the National Science Foundation (NSF) Graduate Research Fellowship under grant no. 1650114.

We thank Kaitlin Kratter for fruitful discussions and comments. We thank the anonymous referee for constructive comments that improved the quality of our work.

This work made use of the \texttt{REBOUND} code which is freely available at http://github.com/hannorein/rebound.

This work made use of the \orbitcodename code. The exact version we used is available via the \orbitcodeversion.

This work made use of the \htofcodename code. We used version \htofversion \citep{htof_zenodo}.
}

\bibliographystyle{aasjournal}
\bibliography{refs.bib}

\end{document}